\documentclass[a4paper,11pt]{article}

\usepackage{jinstpub} 

\usepackage{wasysym}
\usepackage{xspace}
\usepackage{xcolor}
\usepackage{color}
\usepackage[normalem]{ulem}

\newcommand{\vanode}{\ensuremath{V_{\textrm{anode}}}\xspace}
\newcommand{\vback}{\ensuremath{V_{\textrm{back}}}\xspace}
\newcommand{\vmesh}{\ensuremath{V_{\textrm{mesh}}}\xspace}

\newcommand{\am}[0]{$^{241}$Am\xspace}
\newcommand{\dr}[0]{$\Delta r$\xspace}
\newcommand{\yield}[0]{\ensuremath{\mathrm{Y}_{\mathrm{ph}}}\xspace}

\newcommand{\registered}{\ensuremath{^{\textrm{\textregistered}}}\xspace}

\newcommand{\ignoreblock}[1]{}

\title{\boldmath First observation of liquid xenon electroluminescence with a Microstrip Plate}

\author[a,b,1]{G. Martinez-Lema,\note{Corresponding author.}}
\author[c]{V. Chepel,}
\author[a,b]{A. Roy,}
\author[a]{and A. Breskin}

\affiliation[a]{Dept. of Astrophysics and Particle Physics, Weizmann Institute of Science, Rehovot, Israel}
\affiliation[b]{Unit of Nuclear Engineering, Ben-Gurion University of the Negev, Beer-Sheva, Israel}
\affiliation[c]{LIP-Coimbra and Department of Physics, University of Coimbra, 3004-516 Coimbra, Portugal}

\emailAdd{gonzalo.martinez.lema@weizmann.ac.il}

\abstract{We report on the first observation of electroluminescence amplification with a Microstrip Plate immersed in liquid xenon.
The electroluminescence of the liquid, induced by alpha-particles, was observed in an intense non-uniform electric field in the vicinity of 8-$\mu$m narrow anode strips interlaced with wider cathode ones, deposited on the same side of a glass substrate.
The electroluminescence yield in the liquid reached a value of $(35.5 \pm 2.6)$~VUV photons/electron. We propose ways of enhancing this response with more appropriate microstructures towards their potential incorporation as sensing elements in single-phase noble-liquid detectors.}

\keywords{
\\ Noble liquid detectors
\\ Micropattern gaseous detectors
\\ Charge transport, multiplication and electroluminescence in rare gases and liquids
\\ Time Projection Chambers
\\ Dark Matter detectors (WIMPs, axions, etc.)
}

\begin{document}
\maketitle
\flushbottom

\section{Introduction}
\label{sec:intro}

Noble-liquid detectors have been playing a major role in physics experiments and in various other applications, as reviewed in \cite{AprileDoke:2010, Chepel:2013, Majumdar_2021, Akimov:2021book, Aalbers_2023}.
Massive single-phase (liquid), or dual-phase (liquid and gas) noble-gas (Ar and Xe) Time Projection Chambers have become the leading instruments in dark-matter searches, and neutrino experiments.
In these devices, 3D localization is attained through the detection of both primary scintillation light and ionization charges.
For large energy depositions, the latter can be measured as electric current with low-noise preamplifiers. If the ionization charge is too small to be measured directly, it still can be detected through an amplification process --- charge multiplication or electroluminescence. However, both require extremely high electric fields ($\sim 10^5 - 10^6$~V/cm) to develop in the liquid.
Fortunately, for most of the noble gases it is possible to extract the free electrons from the liquid phase to the gas where  amplification becomes possible at orders or magnitude lower electric fields.
This method has given rise to double-phase electron emission argon and xenon detectors \cite{Dolgoshein_1970} --- at present one of the most powerful classes of instruments in dark-matter experiments searching for Weakly Interacting Massive Particles (WIMPs) and for detection of neutrino-nucleus scattering (see already mentioned \cite{AprileDoke:2010, Chepel:2013, Majumdar_2021, Akimov:2021book, Aalbers_2023} and references therein).
The amplification of small primary ionization signals is achieved through electroluminescence produced by the extracted electrons in gas.
The target mass of those detectors is already in the tens of tonnes scale and even bigger detectors are required in order to meet the demand for higher sensitivity to rare events \cite{Aalbers_2023, Liu:2017drf, DarkSide-20k:2017zyg}.

Scaling up of dual-phase detectors, may encounter some difficulties in deploying large-area wire or mesh charge-extracting electrodes.
These include electroluminescence-gap variations (due to electrostatic electrode-wire sagging and staggering, surface ripples, etc.), electrical instabilities at the liquid-gas interface (e.g. spontaneous electron emission) and electron-extraction inefficiencies, which may seriously affect the performance of these detectors.

Thus, the interest in alternative solutions has revived within the dual-phase but also the single-phase (liquid) concepts, many of the proposals relying on the electroluminescence mechanism of amplification.
In the present work, we focus on electroluminescence of liquid xenon (LXe), i.e. single-phase detector variants.
For dual-phase alternatives the reader is referred to \cite{Buzulutskov_2012, Erdal:2018, Buzulutskov_2020, Tesi:2021, Chepel_2023, TesiRPWELL}.
First observation of electroluminescence in LXe has been reported back in 1967 \cite{Dolgoshein:1967}.
Several studies have been undertaken in the upcoming years aiming at development of a particle detector using electroluminescence near a thin (a few $\mu$m) wire \cite{Miyajima:1979, MASUDA1979247, DOKE198287}.
Although good results have been obtained with small prototypes, the wire fragility and shadowing effect represented serious technical challenges for a practical detector.
This line of research has been retaken more recently \cite{Aprile:2014ELthreshold, Kuger:2022, Juyal_2021, Qi:2023}.
Both single wires of 10~$\mu$m and multi-wire configurations have been considered \cite{Juyal_2021}.
A number of novel designs using micropattern structures have been proposed in \cite{Breskin:2022novel}.

For a 10 $\mu$m anode wire, an electroluminescence yield of $\sim$290 photons/electron has been measured at an anode voltage of 6.75~kV~\cite{Aprile:2014ELthreshold}.
The anode wire was placed in between two multi-wire cathode planes placed 8 mm apart.
In a cylindrical geometry with a single anode wire of the same diameter and cathode wires to form a cylinder of 2.5~cm radius, a yield of $\sim$17 photons/electron has been estimated from the data measured at an anode voltage of 3.6~kV \cite{Qi:2023}.
These data are in tension with a previously reported much lower yield in similar conditions \cite{MASUDA1979247}.

Electroluminescence in a non-uniform field near thin wires is accompanied by charge multiplication as the field changes very rapidly with the distance from the wire surface.
Indeed, in practically all referred works some charge multiplication has been observed.
The threshold field for electroluminescence in liquid xenon was estimated to be $\sim$400~kV/cm in \cite{Aprile:2014ELthreshold}.
This is in reasonable agreement with \cite{MASUDA1979247} ($\sim$490 kV/cm), but diverge significantly from \cite{Dolgoshein:1967}, where they report a threshold of $\sim$100~kV/cm in a uniform field.
The charge multiplication threshold was measured in \cite{Aprile:2014ELthreshold} to be $\sim$~700~kV/cm.

Due to the wire fragility and mechanical instability under the high voltage required for efficient production of secondary photons ($\sim$5~kV), it has been recently proposed \cite{Breskin:2022novel} to use narrow metallic strips deposited on a dielectric support (e.g. quartz) instead of thin wires.
Such micro-pattern structures are well known in gaseous detectors being the MicroStrip Plate (MSP) one of them \cite{OED1988351}.
It is worth noting that charge multiplication near the 8-$\mu$m wide anode strips of a MSP has been previously observed in liquid xenon \cite{POLICARPO1995568};
a charge gain of $\sim$10 has been reported although the photon emission was not detected in that work.

In this article, we report our results on the study of electroluminescence in LXe near the anode strips of a MSP having exactly the same geometry as in \cite{POLICARPO1995568}.
The current limitations are discussed and further directions are proposed for the development of advanced single-phase detectors of both scintillation photons and ionization electrons.
\section{Experimental setup}
\label{sec:setup}

\begin{figure}
    \centering
    \includegraphics[height=50mm]{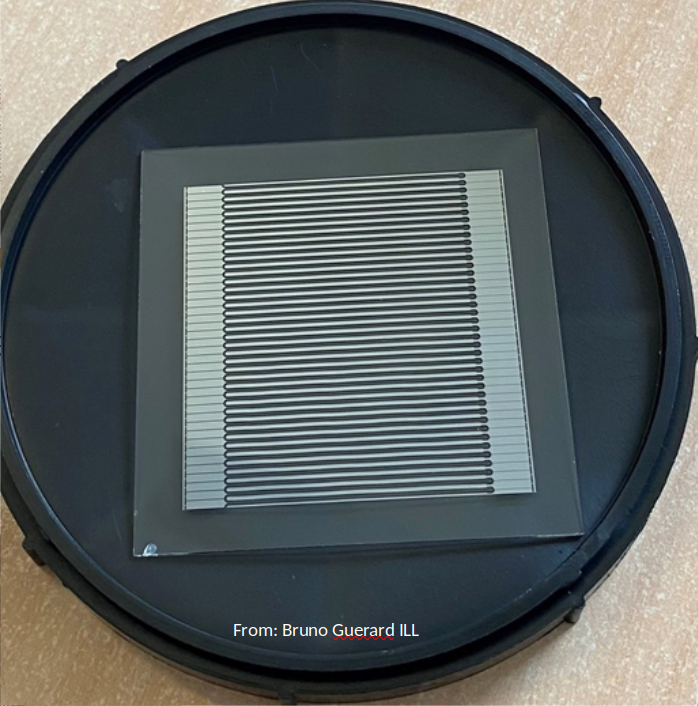}
    \includegraphics[height=50mm]{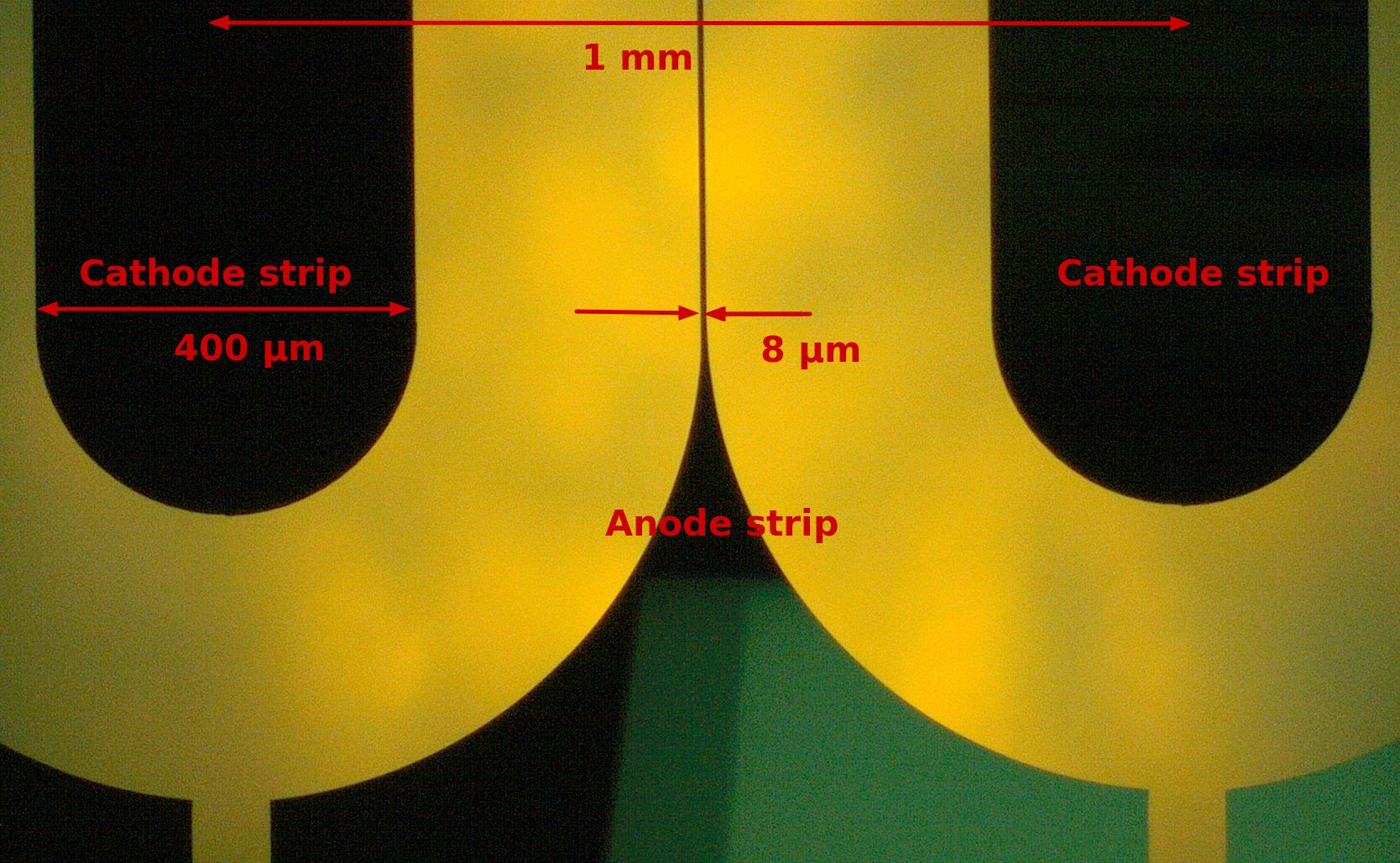}
    \caption{Left: picture of the ILL-6C microstrip plate used in the experiments. Right: microscope picture of the plate.}
    \label{fig:plate}
\end{figure}

The experiments were carried out with a ILL-6C microstrip plate described in~\cite{OED199534}, shown in Figure~\ref{fig:plate}.
The anode strips, 8~$\mu$m wide, alternate with 400~$\mu$m wide cathode ones with an anode-to-anode pitch of 1~mm.
The strips are made of 150~nm thick chromium deposited on a 0.55~mm thick D263 Schott glass substrate without a backplane electrode.
The latter was added in our experiments by mechanically pressing a 100~$\mu$m thin stainless steel foil against the glass substrate.
All anode strips and cathode ones are interconnected on opposite sides of the plate (Figure~\ref{fig:plate}).
The plate dimensions are 40 x 40 mm$^2$ with an active area of 28 x 28~mm$^2$.

\begin{figure}
    \centering
    \includegraphics[width=0.75\textwidth]{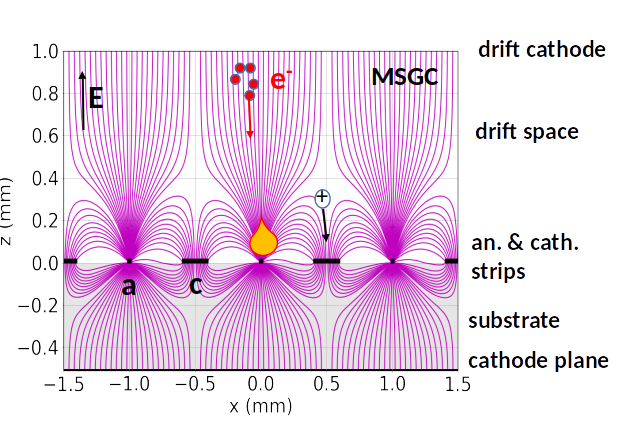}
    \caption{Schematic representation of the operating principle of an MSGC. Electrons released in the drift volume drift towards the thin anode strips where they induce electroluminescence and charge avalanche under the high electric fields in the vicinity of the electrode. Ions released in the avalanche drift to the neighboring cathode strips. From \cite{Breskin:2022novel}.
    }
    \label{fig:principle}
\end{figure}

The electric field map and the operation principle of the MSP-based MicroStrip Gas Chamber (MSGC), originally designed for multiplication in gas,  are depicted in Figure~\ref{fig:principle}.

\begin{figure}
    \centering
    \includegraphics[width=\textwidth]{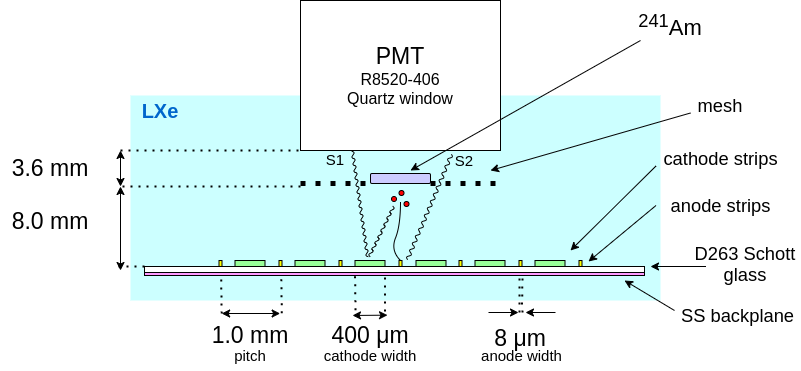}
    \caption{Schematic drawing of the setup used. An \am alpha source with a diameter of 6 mm (active area <2~mm) is placed at the center of a woven mesh.
    Behind the mesh, a 1-inch square VUV-sensitive PMT collects the primary scintillation light reflected off the metallic components and the electroluminescence light produced in the vicinity of the MSP anode strips, located 8~mm below the source-supporting mesh.
    The back of the plate consists of a 100-$\mu$m-thick stainless steel foil that can be electrically biased.
    During experiments, the source and the mesh were biased to -2~kV, and the cathode strips were grounded.
    The anode strip bias voltage was varied between 0 and 2~kV and the backplane bias voltage between 0 and -3~kV depending on the measurement.
    The PMT was operated at -750 V.
    }
    \label{fig:setup_scheme}
\end{figure}

\begin{figure}
    \centering
    \includegraphics[width=0.5\textwidth]{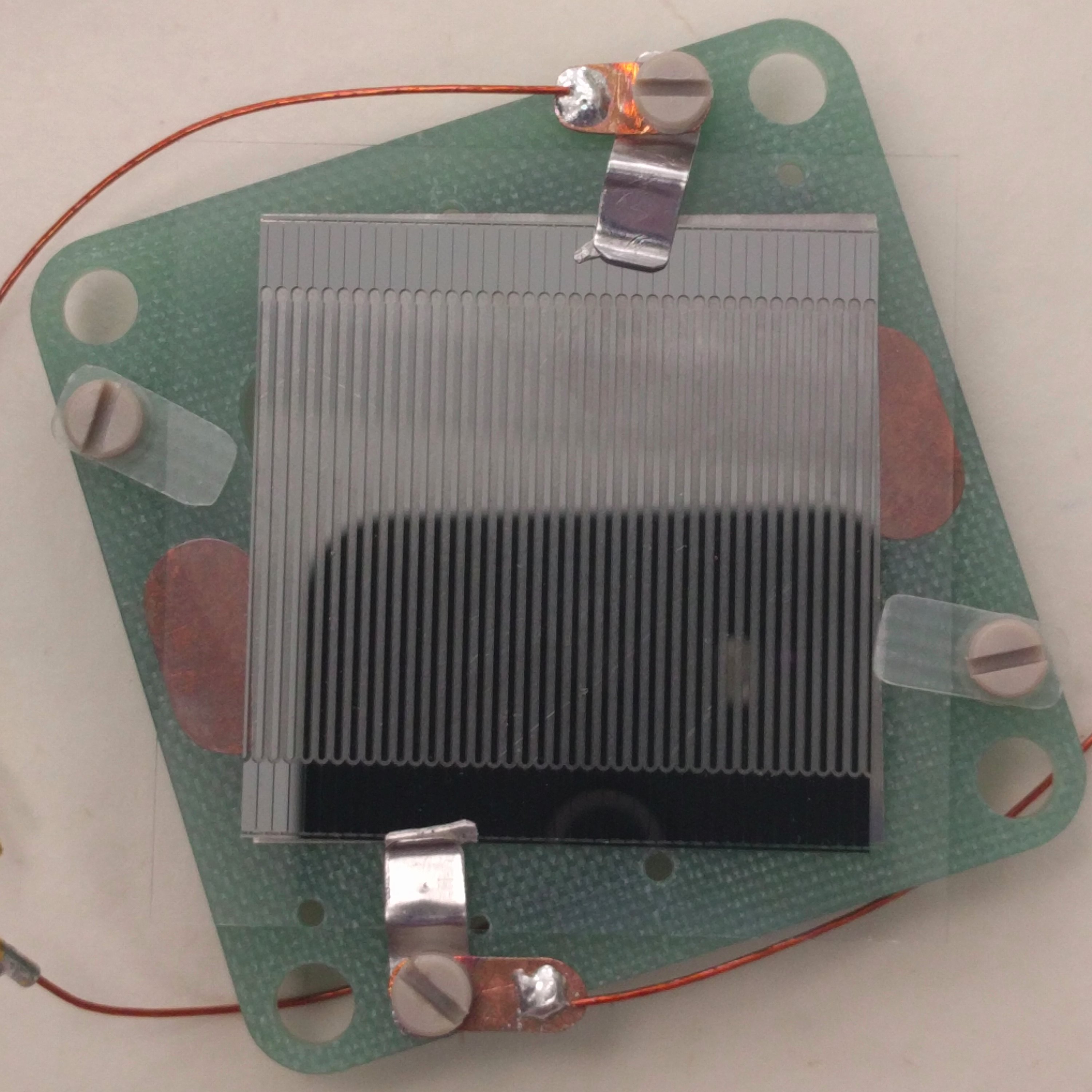}
    \caption{Assembly of the setup.}
    \label{fig:setup_picture}
\end{figure}

The detector assembly is depicted schematically in Figure \ref{fig:setup_scheme}.
The MSP plate sits on a holder made of FR4.
Electrical connections to the anode and cathode strips are achieved using indium pressed against the interconnected areas of the strips as shown in Figure \ref{fig:setup_picture}.
Above the MSP, a 1x1-inch$^2$ VUV-sensitive PMT (Hamamatsu R8520-406 with a quartz window) is housed in a PTFE-made holder.
A 5.5~MeV \am alpha source ($\sim$40 Bq) with an active area of <2~mm in diameter, deposited on a stainless steel substrate, is pointed at the MSP.
The source is located at the center of a woven stainless-steel mesh (50~$\mu$m diameter wire; 0.5~mm opening) with an optical transparency of 81\%, soldered on a 1.6~mm thick FR4 support.
The mesh frame is in turn attached to the PMT holder using PEEK screws.
The MSP and PMT holders are joined together using PEEK rods and spacers.
The total distance between the MSP plate and the PMT is 11.6 mm; the resulting PMT solid angle (defined by the PMT area and the shadow of the alpha source as seen in Figure \ref{fig:setup_scheme}) is $\sim$1.3 sr (10.5\%).
The entire system is suspended from the top flange of the cryostat.

The experiments were carried out in the MiniX cryostat described in detail elsewhere \cite{Erdal:2015kxa}.
The chamber was filled with liquid xenon (LXe) so that the PMT window fully immersed in the liquid.
The LXe was kept at 180~K during the experiments, corresponding to a vapor pressure of 2.15 bar.
Before starting xenon condensation, the chamber was pumped down to a vacuum $\lesssim 10^{-5}$~mbar for at least 8 hours.
The Xe gas circulated in the system through a SAES hot getter (model PS3-MT3-R-2) at a rate of $\sim$1.2 standard liters per minute (slpm) for several days before taking data.

CAEN N471A power supplies were used to apply voltages to the electrodes.
The PMT waveforms were digitized and recorded with a Tektronix 5204B oscilloscope.
The waveform analysis was performed offline.

\section{Results}
\label{sec:results}

The measurements presented here were obtained by recording PMT waveforms under different voltage configurations of the MSP.
We will use the following notation: \vanode for the bias applied to the anode strips and \vback for the voltage applied to the backplane.
The mesh holding the alpha source was biased at $\vmesh = -2$~kV and the cathode strips were grounded for all the measurements reported.
The PMT voltage was kept at -750~V.

\begin{figure}
    \centering
    \includegraphics[width=\textwidth]{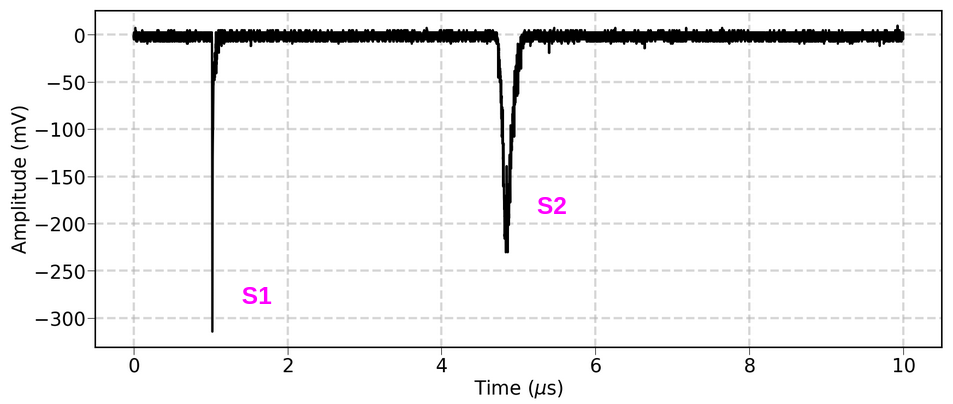}
    \caption{A typical PMT waveform obtained for an $\alpha$-particle interaction in LXe, with the primary scintillation (S1) and electroluminescence (S2) signals labeled. The waveform was recorded at $\vanode~=~1.6$~kV, $\vback = -2$~kV, grounded cathode strips and $\vmesh =    -2$~kV. The oscilloscope was triggered by the S1 pulse.}
    \label{fig:waveform}
\end{figure}

Figure \ref{fig:waveform} shows a typical waveform obtained for an alpha-particle interaction in LXe.
The primary scintillation light produced is reflected off the metallic surface of the strips and collected on the PMT (S1).
The ionization electrons drift towards the MSP where the intense electric field in the vicinity of the anode strips is sufficient to induce electroluminescence, which is detected by the PMT (S2).

Most of the S2 light is detected directly by the PMT (partially masked by the source), with a small fraction of it resulting from reflections.
Since light generation occurs very close to the strips (see Section \ref{sec:simulations}), part of the light impinging on the strips is reflected back towards the PMT.
This secondary contribution is given by the reflectivity of chromium at 178 nm.

\subsection{Dependence on anode-strip bias voltage}
\label{sec:vanode}

\begin{figure}
    \centering
    \includegraphics[width=0.49\textwidth]{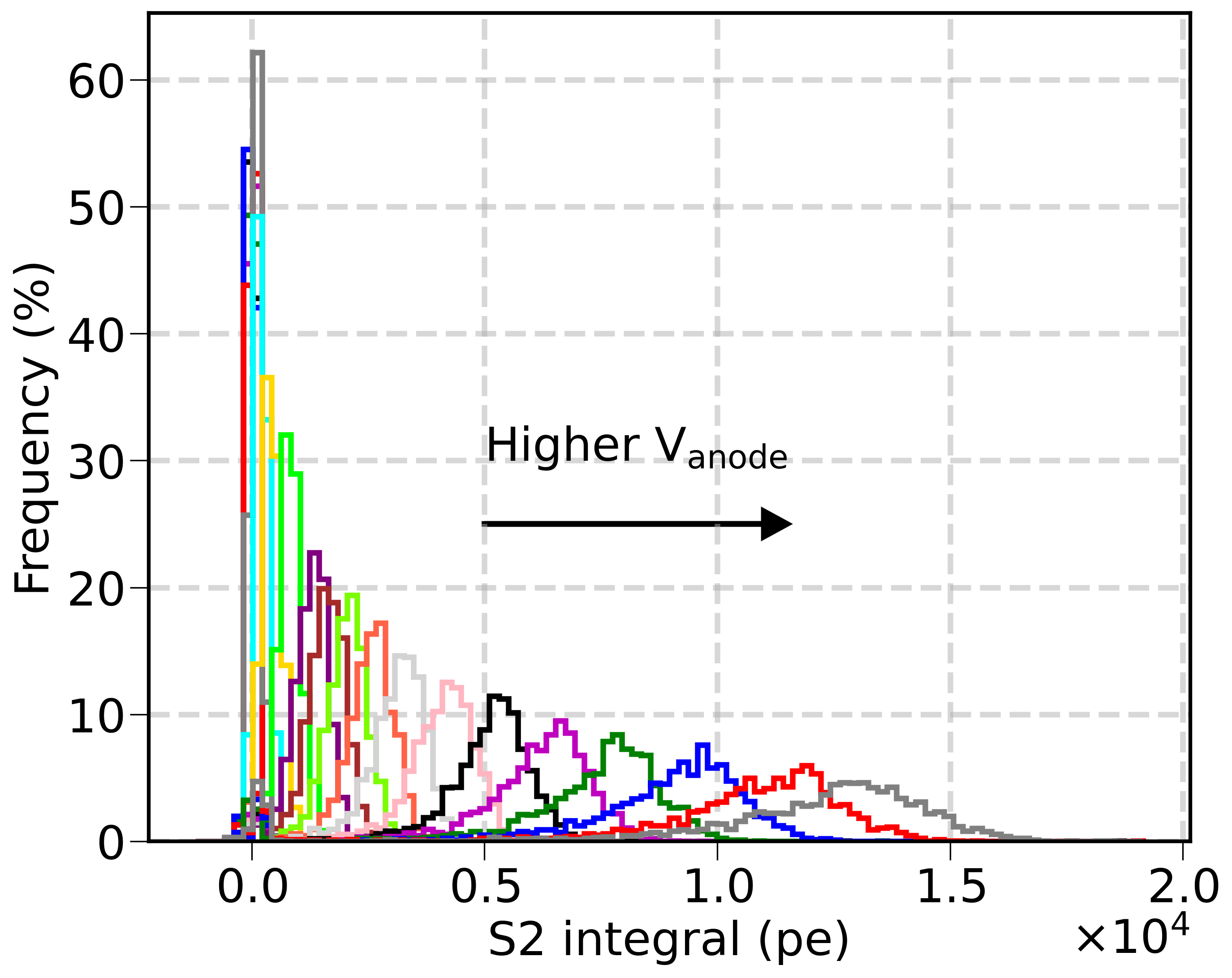}
    \includegraphics[width=0.49\textwidth]{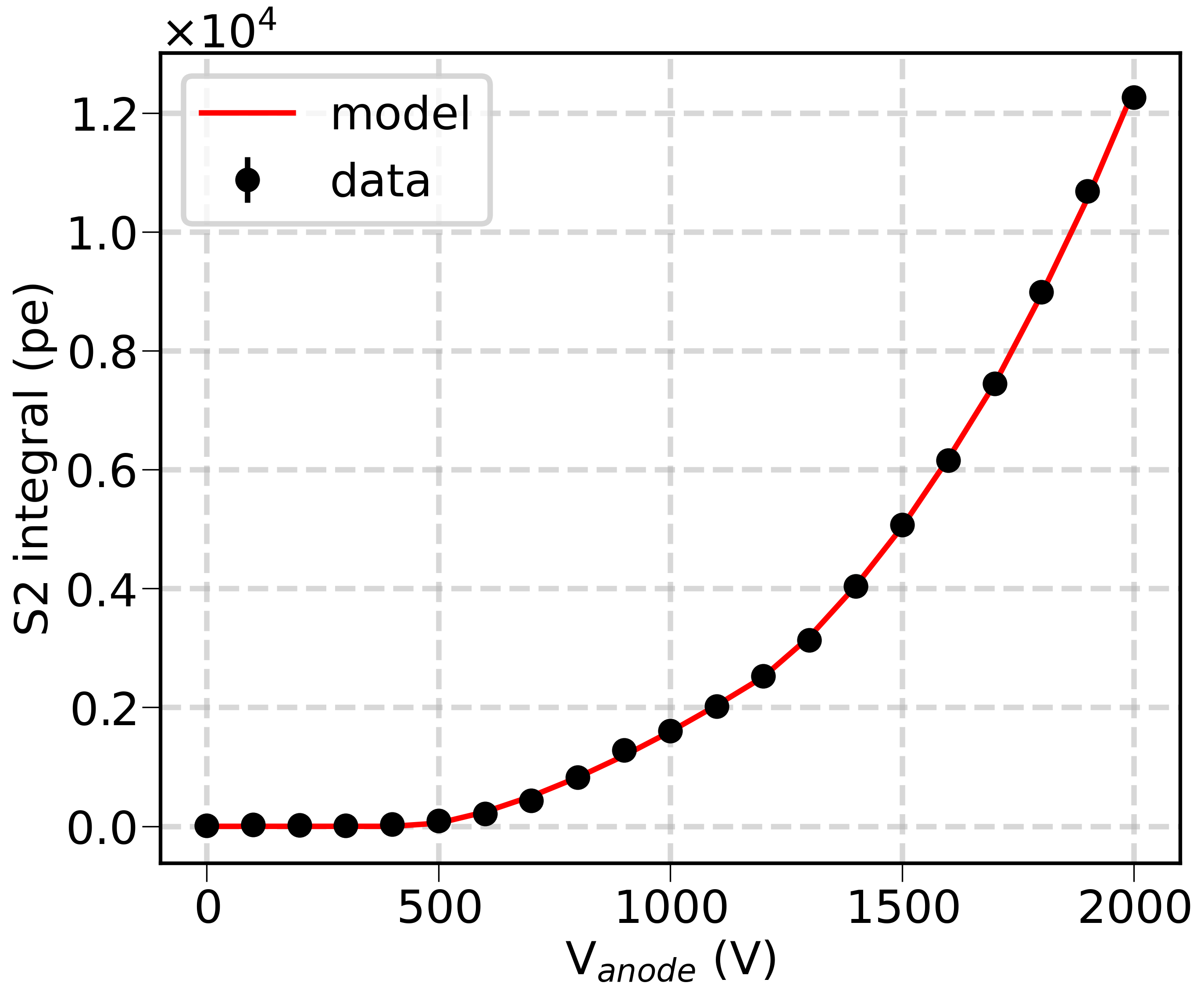}
    \caption{Left: S2 integral spectra as a function of \vanode in the range 500 V to 2000 V.
    Right: Mean value of the integrated S2 signal as a function of the voltage on the anode strips.
    The statistical uncertainties are too small to be shown.
    The data were taken at $\vback = -2$~kV, grounded cathodes and $\vmesh = -2$~kV.
    }
    \label{fig:vanode_scan}
\end{figure}

For the first part of the analysis, PMT waveforms were acquired at $\vback = -2$~kV and for values of \vanode up to 2 kV.
At this point, discharges between anode and cathode strips were observed, which prevented reaching higher anode-strip voltages.
For each voltage, $\sim$5000 waveforms were recorded;
for each waveform, the S2 signal was integrated over a fixed time window (4.5 to 5.5 $\mu$s in Figure \ref{fig:waveform}).
The S2 integrals were translated into number of photoelectrons (pe) by an in-situ PMT calibration using the method described in \cite{DOSSI2000623}.
The left panel of Figure \ref{fig:vanode_scan} shows the S2-integral spectra for the different values of \vanode.
The right panel of Figure \ref{fig:vanode_scan} depicts the mean value of the integrated S2 signal as a function of the voltage on the anode strips. 
We see that the S2 signals first appear at $\vanode \sim 500$~V.

The mean value of the S2 spectra is then used to measure the relative light yield.
For \vanode~=~2~kV, we measure $1.24 \cdot 10^4$ pe.
The solid angle in this setup is $\Omega = 10.5$\%, and the transparency of the mesh is $T = 81$\%.
We assume a PMT QE of 28\% \cite{Hamamatsu} and a chromium reflectivity of $r = (10.5 \pm 0.6)$\% (see estimate in Appendix \ref{app:reflectivity}).
The energy of the $\alpha-$particles from the film-coated source used in our setup was previously measured to be $(4.7 \pm 0.2)$~MeV.
Assuming a $W-$value for ionization of liquid xenon of $W_i = (15.6 \pm 0.3)$~eV \cite{PhysRevA.12.1771} and that $(4.5 \pm 0.2)$\% electrons escape recombination~\cite{APRILE1991119}, we expect $N_{ie} = (1.33 \pm 0.06) \cdot 10^4$ ionization electrons (ie) to be extracted from the track of each alpha-particle.
Under these assumptions, the light yield measured at $\vanode = 2$~kV is $\yield = (35.5 \pm 2.6)$~photons per electron reaching the amplification region near the MSP plate.

The right panel of Figure \ref{fig:vanode_scan} displays the mean value of the S2 spectra as a function of \vanode.
The EL and CM model suggested in \cite{Aprile:2014ELthreshold} and shown in Eqs. \ref{eq:model} is fitted (solid red line) to the data (black circles).

\begin{equation}
    \begin{split}
        \frac{dN_e     }{dx} &= N_e(x) \cdot \theta_0 \cdot \exp{- \frac{\theta_1}{E(x) - \theta_2}} \\
        \frac{dN_{ph}}{dx} &= N_e(x) \cdot \theta_3 \cdot \left( E(x) - \theta_4 \right)
    \end{split}
    \label{eq:model}
\end{equation}

\noindent
where $N_e(x)$ and $N_{ph}(x)$ are the number of electrons and photons, respectively, at a given position, $E(x)$ is the local electric field obtained from finite-element calculations performed in COMSOL Multiphysics\registered and $\theta_k$ are adjustable parameters.
$\theta_2$ and $\theta_4$ can be interpreted as the thresholds for charge multiplication and electroluminescence, respectively.

From this fit, we obtain $\theta_2$~=~785~kV/cm and $\theta_4$~=~465~kV/cm.
These numbers are in reasonable agreement with those presented in \cite{Aprile:2014ELthreshold} where values of $\theta_2$~=~725~kV/cm and $\theta_4$~=~412~kV/cm are reported.
Based on this model, we estimate that the charge multiplication factor in our configuration was approximately 3 for \vanode~=~2~kV.
This value is approximately 3-fold lower than that obtained in \cite{POLICARPO1995568}.
In Section \ref{sec:discussion}, we propose some explanation for this discrepancy.

\subsection{Dependence on backplane bias voltage}
\label{sec:vback}

\begin{figure}
    \centering
    \includegraphics[width=0.5\textwidth]{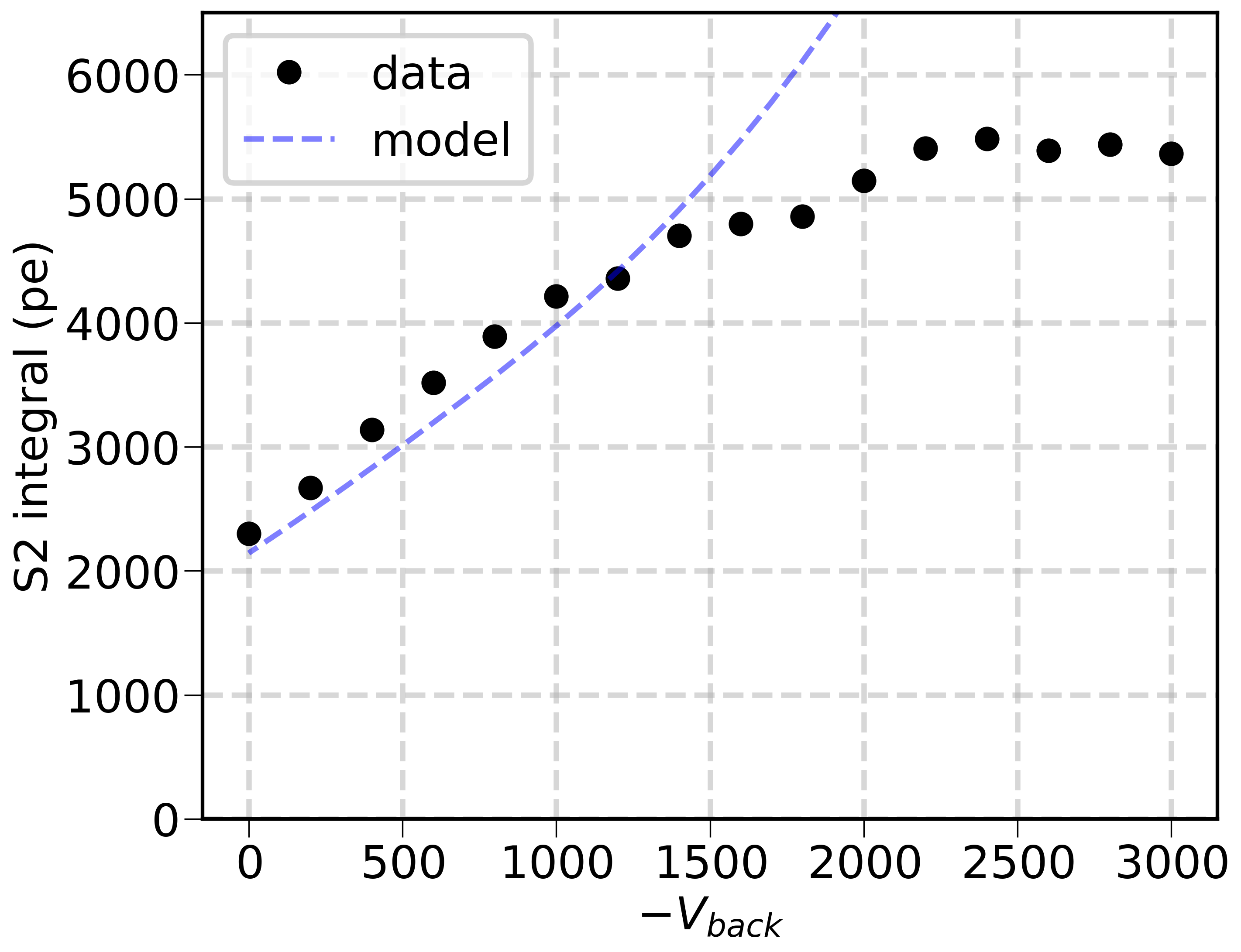}
    \caption{Mean value of the integrated S2 signal as a function of the backplane voltage (black points) and trend (dashed blue line) predicted by the model from Eqs. \ref{eq:model}.
    The data were taken at \vanode~=~1.6~kV, grounded cathodes and $\vmesh = -2$~kV.
    The trend on the graph uses an arbitrary scale for visual comparison only.
    } 
    \label{fig:vback_scan}
\end{figure}

In this study, PMT waveforms were acquired with the anode strips at \vanode~=~1.6~kV, with values of \vback ranging from 0 to -3~kV.
For each voltage, $\sim$5000 waveforms were recorded and the waveform processing was the same as the one described in Section \ref{sec:vanode}.
Figure \ref{fig:vback_scan} shows the mean value of the S2 integral spectra (black points) as a function of \vback.
We observe a strong increase of S2 between 0 and -2 kV followed by a plateau.
The dashed blue line represents the predicted trend based on COMSOL calculations and on the model from Eqs. \ref{eq:model}.
This uses an arbitrary scale and it is shown for visual comparison only, not representing real predictions.
Possible reasons for this discrepancy are discussed in Section \ref{sec:discussion}.

\subsection{Energy resolution}

\begin{figure}
    \centering
    \includegraphics[width=0.49\textwidth]{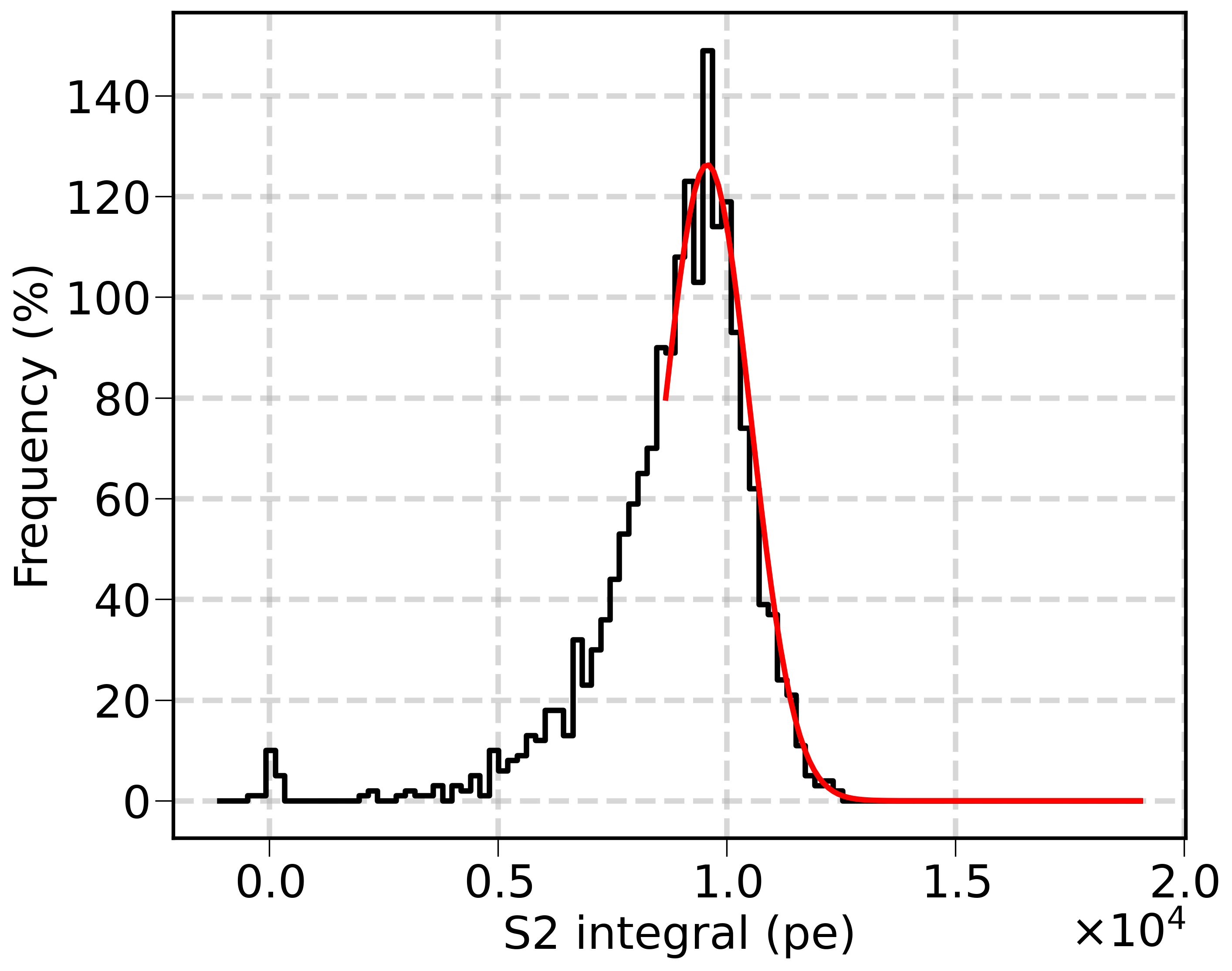}
    \includegraphics[width=0.49\textwidth]{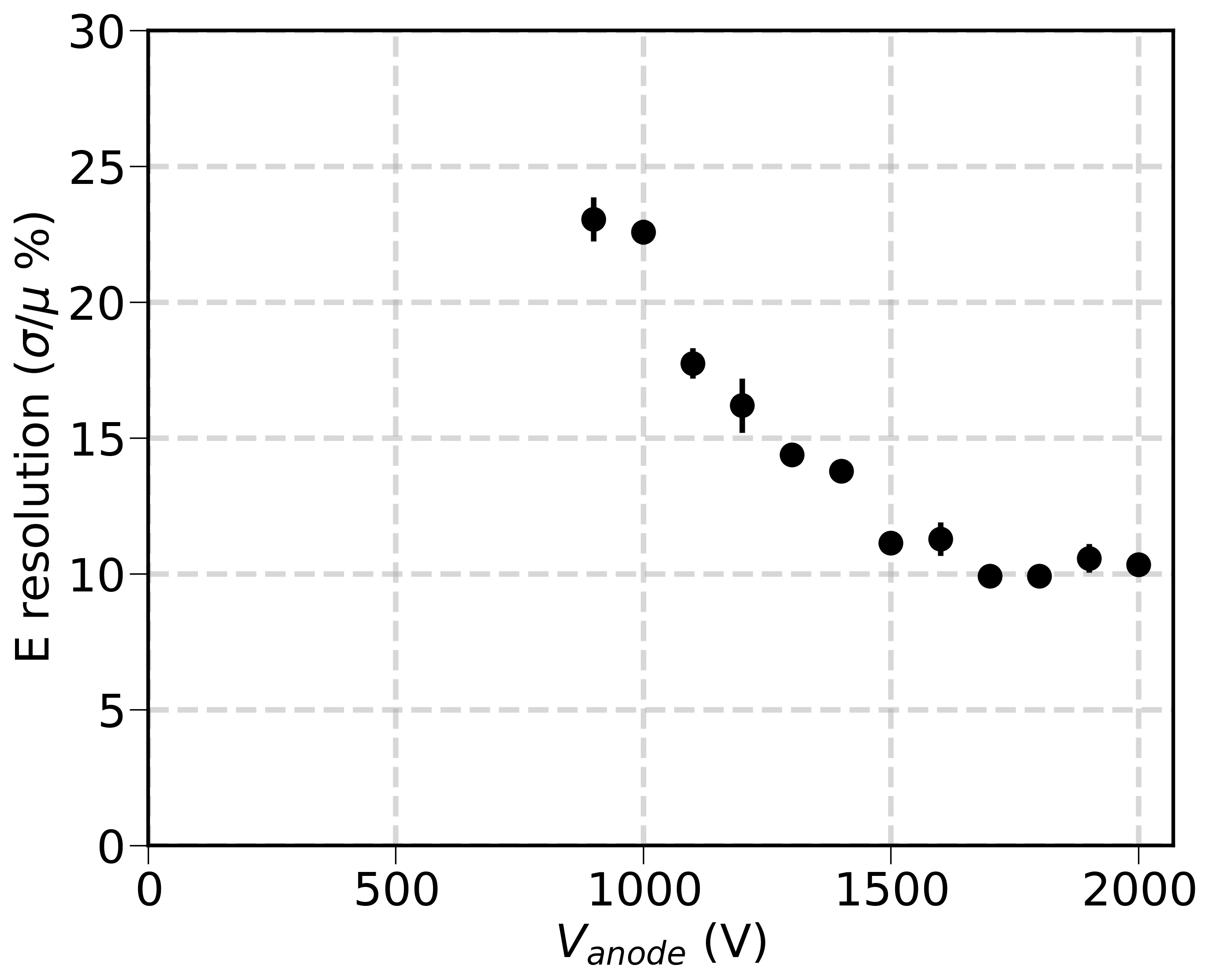}
    \caption{Left: S2 energy spectrum (black) measured at \vanode~=~1.8~kV and \vback~=~-2~kV. and fit of a gaussian model to the high-energy side (red). Right: energy resolution as a function of \vanode.}
    \label{fig:e_res}
\end{figure}

The same dataset from Section \ref{sec:vanode} was analyzed to obtain the energy resolution of the MSP for alpha particles.
A gaussian function was fitted to the high-energy side of the energy spectrum (the left side contains contributions from partial tracks) for each value of \vanode.
The left panel of Figure \ref{fig:e_res} shows the spectrum for \vanode~=~1.8~kV (black line) with the gaussian fit superimposed (red line).
The energy resolution, defined as $\sigma/\mu$, where $\sigma$ and $\mu$ are the standard deviation and mean of the gaussian distribution, is shown in the right panel of Figure \ref{fig:e_res}.
The best energy resolution, $\sigma/\mu = 9.9\%$, was obtained at $\vanode = 1.8$~kV.
This result is similar to that obtained with thin anode wires in \cite{Aprile:2014ELthreshold} for a similar anode voltage, once we account for the difference in photoelectron statistics (see Section \ref{sec:discussion}).

\section{Extrapolation to other MSP geometries}
\label{sec:simulations}

The performance of the MSGC is limited by anode-to-cathode strip discharges due to the proximity of the strips laying on the same surface.
Therefore, some authors have proposed different MSP geometries that could achieve better performance by increasing the voltage limit imposed by discharges.
Here we consider two alternative designs proposed for single-phase noble-liquid detectors \cite{Breskin:2022novel}: the Coated-Cathode Conductive Layer electrode (COCA-COLA) \cite{BOUCLIER199174} and the Virtual Cathode Chamber (VCC) \cite{CAPEANS199717}.

\begin{figure}
    \centering
    \includegraphics[width=0.85\textwidth]{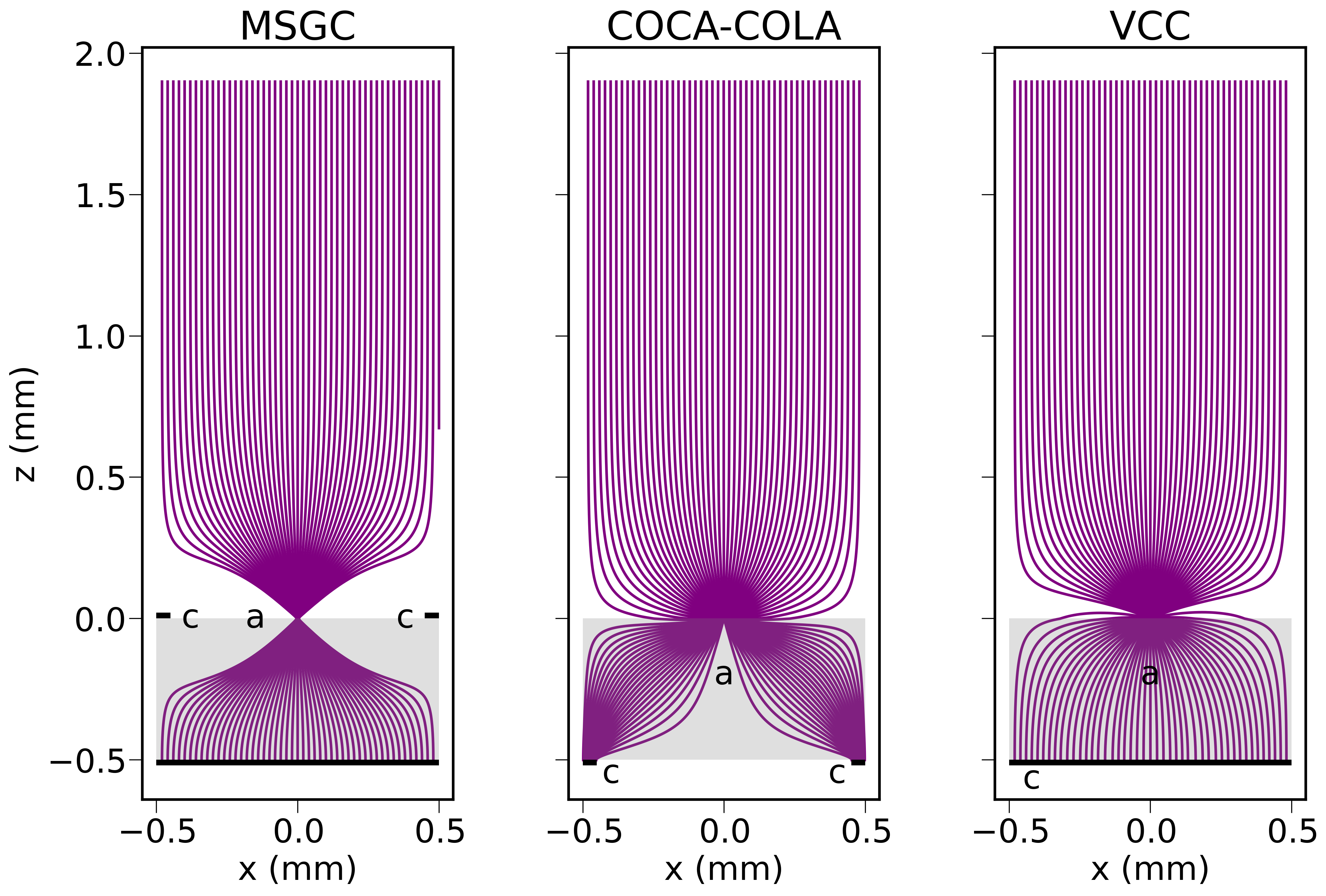}
    \caption{Typical electric field lines for the MSGC (left), COCA-COLA (center) and VCC (right) geometries. The presence of the cathode strips in the MSGC configuration introduces a focusing effect of the field lines onto the anode strip.}
    \label{fig:field_configurations}
\end{figure}

Figure \ref{fig:field_configurations} shows a comparison of the field lines between MSGC, COCA-COLA and VCC obtained in COMSOL Multiphysics\registered.
Unlike the MSGC, the COCA-COLA and VCC designs place the cathode electrodes on the back of the plate.
In the case of COCA-COLA, the cathode consists of narrow ($\sim$100 $\mu$m) strips while the VCC has a fully conductive surface or a grid structure.
Since the cathode and anode electrodes are physically separated by an insulator, the discharge limit is expected to be much higher, allowing for operation at considerably higher voltages.
Here we assume that the cathode electrode is grounded and that an anode voltage of 5 kV is feasible.

\begin{figure}
    \centering
    \includegraphics[width=0.7\textwidth]{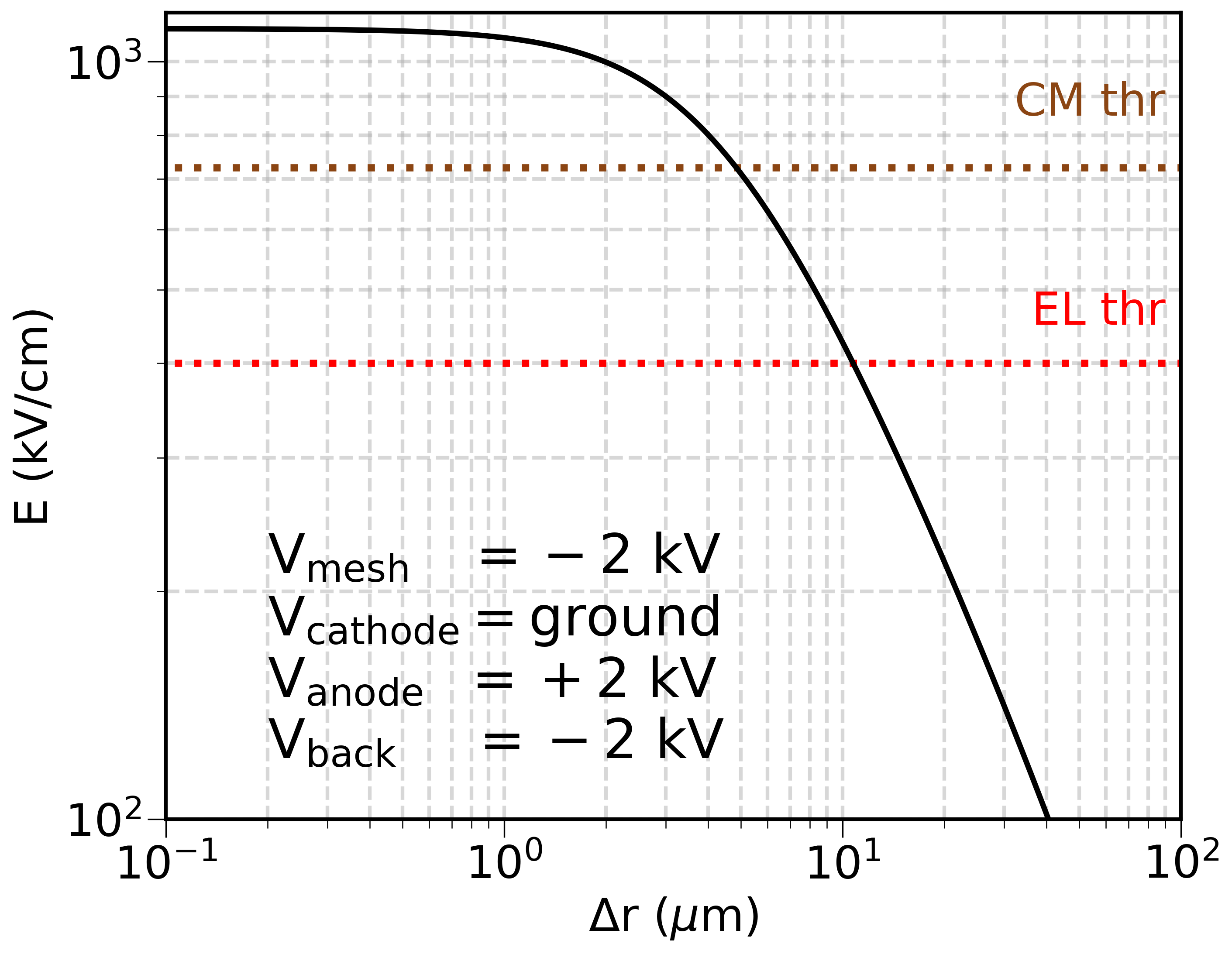}
    \caption{Electric field as a function of distance to the center of the anode strip along the vertical drift line for the MSGC used in our experiment and described in Section \ref{sec:setup} (solid black line). The calculated data correspond to the discharge-limited $\vanode = 2$~kV and to $\vback = -2$~kV. The red and brown dotted lines correspond to the EL and CM thresholds obtained in this work.}
    \label{fig:msgc_field_dr}
\end{figure}

The finite-element calculations provide the field strength in a fine mesh of points in space and the field lines, which are taken as a proxy to the electron trajectories.
Using these two outputs, we can predict the electric field seen by an electron following a drift line.
This is shown in Figure~\ref{fig:msgc_field_dr}, where we display the electric field as a function of \dr, the distance to the surface of the anode strip measured along the drift line, for the MSGC described in this article.
The plot is for the shortest field line, i.e. the one starting at the center of the anode strip.
The EL and CM thresholds, obtained in Section \ref{sec:vanode}, are also shown for reference.
Based on this simulation, we estimate that, for \vanode~=~2~kV, the EL and CM thresholds are crossed at a distance of ${\sim}10~\mu$m and ${\sim}5~\mu$m from the anode, respectively.

\begin{figure}
    \centering
    \includegraphics[width=0.7\textwidth]{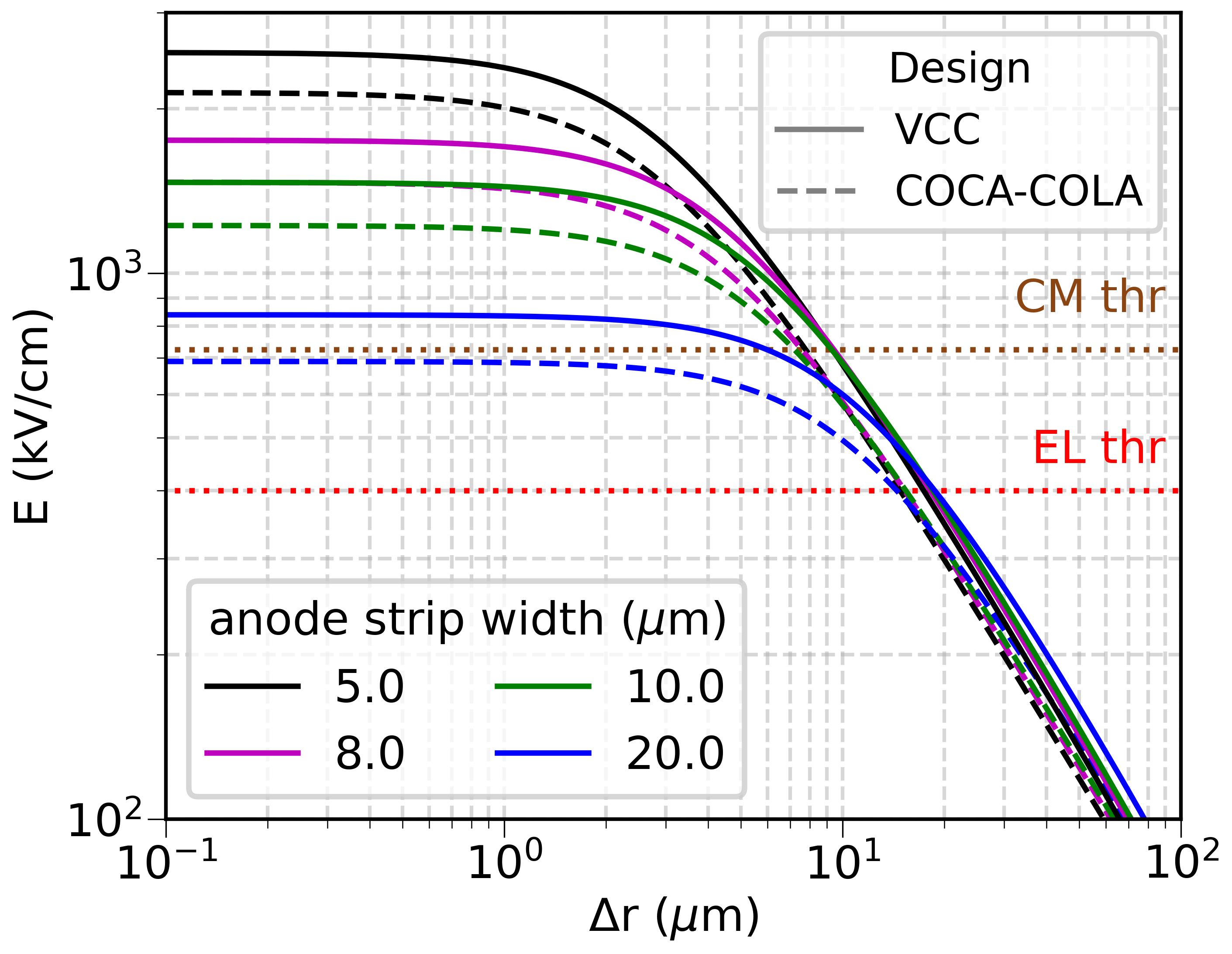}
    \caption{Electric field as a function of distance to the center of the anode strip along the vertical drift line for the COCA-COLA (dashed lines) and VCC (solid lines) geometries for various anode-strip widths. The red and brown dotted lines correspond to the EL and CM thresholds. The voltage configuration is \vanode~=~5~kV, \vmesh~=~-2~kV and grounded cathode plane/strips.}
    \label{fig:coca_vcc_wanode}
\end{figure}

Figure \ref{fig:coca_vcc_wanode} shows a similar plot for the COCA-COLA and VCC geometries with strip widths of 5 to 20 $\mu$m.
We observe that the VCC configuration produces more intense electric fields than the COCA-COLA for a given anode-strip width and that the electric field strength increases as the strip width decreases.

\begin{figure}
    \centering
    \includegraphics[width=0.7\textwidth]{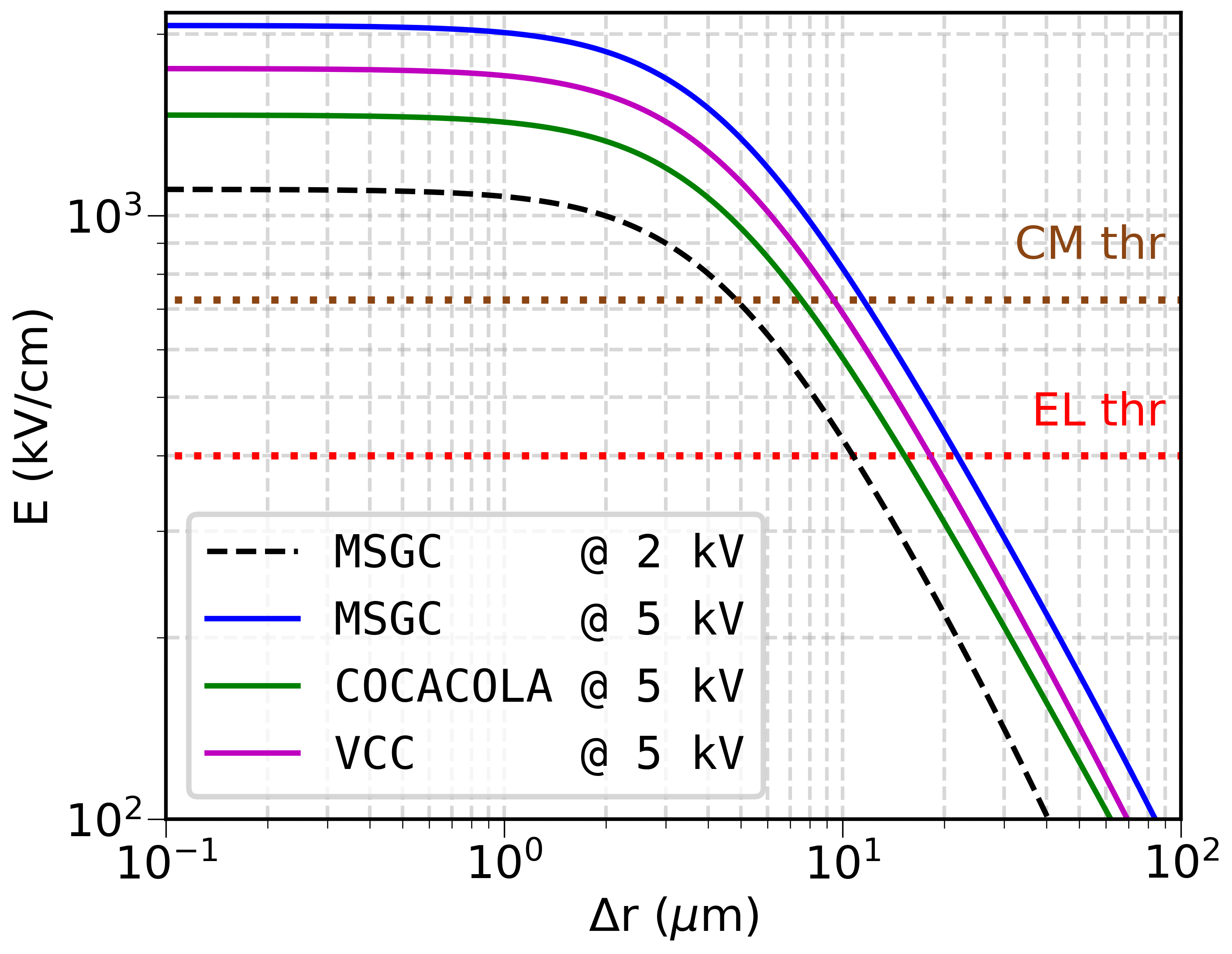}
    \caption{Comparison of the electric field as a function of the distance to the center of the anode strip along the vertical drift line for the three configurations. The red and brown dotted lines correspond to the EL and CM thresholds. The width of the anode strips is 8~$\mu$m, the width of the cathode strips is 200~$\mu$m for the COCA-COLA and 400~$\mu$m for the MSGC. All geometries have a pitch of 1~mm and a 0.55-mm-thick substrate. For the MSGC, \vback=-2~kV. In all cases, \vmesh~=~-2~kV and grounded cathode strips/plane. The dashed black line corresponds to the current experimental condition. The blue line (MSGC~@~5~kV) shows an unrealistic scenario and it is only displayed for comparison.}
    \label{fig:comparison}
\end{figure}

In Figure \ref{fig:comparison} we compare the electric field along the vertical drift line for the highest anode voltage (2~kV) in the current MSGC (dashed black line) to the one calculated for the COCA-COLA (green) and VCC (magenta) geometries at \vanode~=~5~kV.
The values for the MSGC operated at an unfeasible $\vanode = 5$~kV (blue) are also shown for reference.
This graph illustrates how the MSGC would be the optimal geometry if it were not limited by discharges.
Both the COCA-COLA and VCC devices can provide higher light yields than the MSGC operated at \vanode~=~2~kV due to their more intense fields and their extended amplification range.

\begin{figure}
    \centering
    \includegraphics[width=0.49\textwidth]{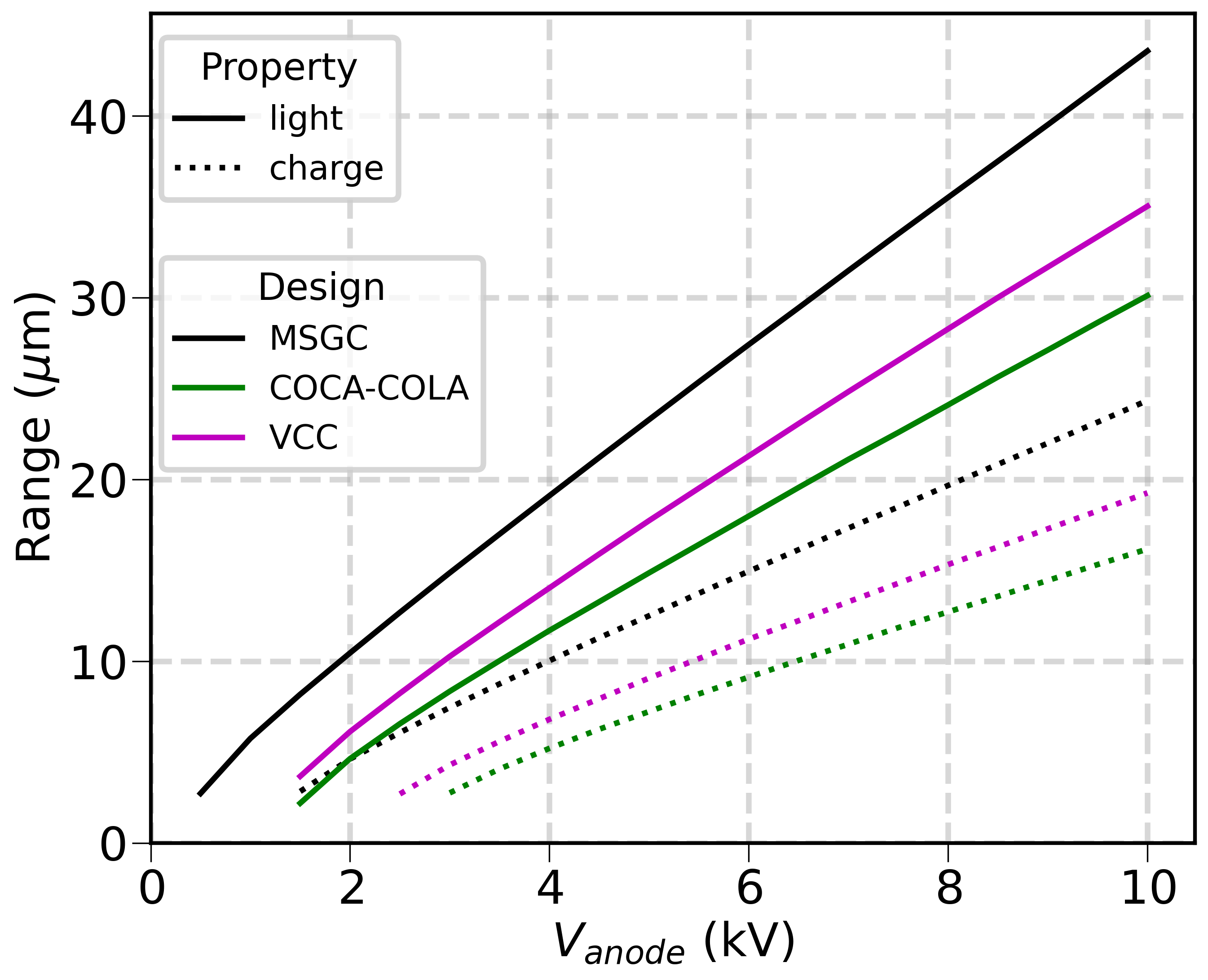}
    \includegraphics[width=0.49\textwidth]{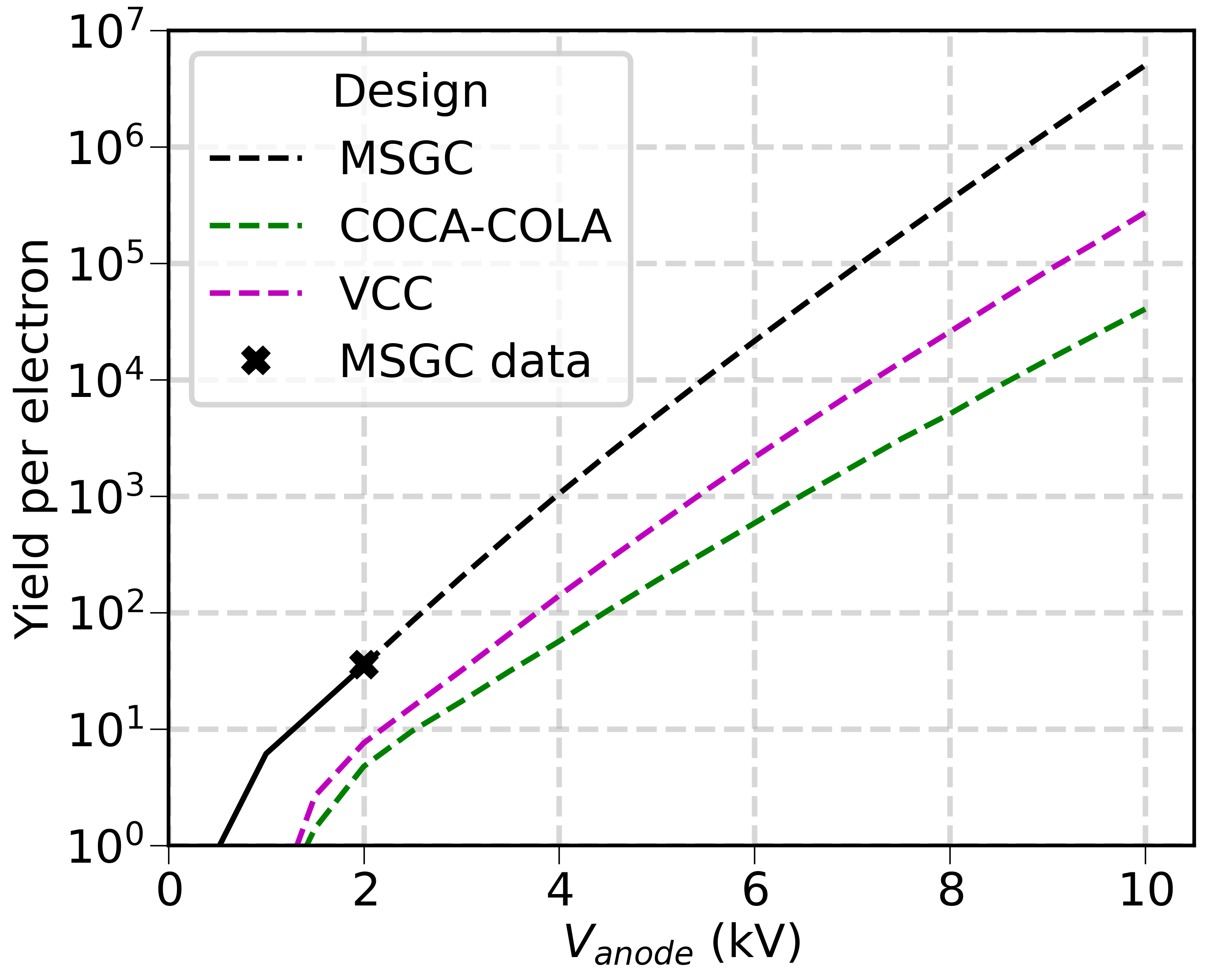}
    \caption{Left: extent of the \dr region along the shortest field line for which the electric field strength is above the EL (solid lines) and CM (dotted lines) thresholds as a function of the anode voltage for the MSGC (black), COCA-COLA (green) and VCC (magenta) geometries.
    Right: light yield predicted by the model in Eqs. \ref{eq:model} defined as number of photons per drifting electron normalized to the value measured in this work.
    The solid line represents the range of voltages for which the yield was measured while the dashed lines represent predictions based on the model.
    All three configurations feature 8-$\mu$m-wide anode strips, 1-mm pitch and a 0.55-mm-thick substrate.
    The cathode-strip width for the COCA-COLA device is 200 $\mu$m.
    In all cases \vmesh~=~-2~kV and the cathode strips/plane are grounded.
    For the MSGC, \vback~=~-2~kV.
    The measured light yield for the MSGC at \vanode~=~2~kV (black cross) is included for reference.
    }
    \label{fig:msgc_coca_vcc_vanode}
\end{figure}

In the left panel of Figure \ref{fig:msgc_coca_vcc_vanode} we show the extent of the \dr region in which the electric field strength is above the electroluminescence (solid lines) and charge multiplication (dotted lines) thresholds for the MSGC (black), COCA-COLA (green) and VCC (magenta) geometries.
The obtained data allow to predict the light and charge yields of these devices by numerical integration of Eqs. \ref{eq:model}.
Using the parameters $\theta_k$ reported in \cite{Aprile:2014ELthreshold} and normalizing the result to the light yield measurement in this work one obtains the electroluminescence yield as a function of \vanode as shown in the right panel of Figure \ref{fig:msgc_coca_vcc_vanode}.
The solid line indicates the range of voltages for which the light yield was measured in this work, and the dashed lines the extrapolations based on the model.
The value of the light yield measured in Section \ref{sec:vanode} for the MSGC at \vanode~=~2~kV (black cross) is also included for reference.
Our calculations show that, at \vanode~=~5~kV, the COCA-COLA device would produce $\yield \sim$171~photons/electron while the VCC device would yield $\yield \sim$545~photons/electron (both structures at $\vmesh = -2$~kV and grounded cathodes).

\section{Summary and discussion}
\label{sec:discussion}

We report on the first observation of electroluminescence amplification with a microstrip plate immersed in LXe.
The data were obtained in LXe with 4.7~MeV alpha particles and a microstrip plate designed for the operation as a microstrip gas chamber electrode.

Limited by the potential applied between the anode and cathode strips of the current plate, we estimated a light yield of $\yield = (35.5 \pm 2.6)$~photons/electron at an anode-to-cathode voltage difference of 2~kV.
This value is approximately 3 times lower than that predicted by the electroluminescence and charge-multiplication model from Eqs. \ref{eq:model} when using the parameters obtained in \cite{Aprile:2014ELthreshold} with a 10-$\mu$m diameter wire in a parallel plate geometry.
This might be an indication that the parameters obtained for a single wire are not applicable to our case.
Another possible explanation for this difference is that the integration of Eqs. \ref{eq:model} is performed along a drift line.
The realistic simulation of the electron propagation in the liquid, might result in a lower light yield.
Moreover, our calculations are based on the shortest field line, while the data consist of the integration over a cloud of electrons following different lines.

The charge gain was estimated to be $\sim$3 at \vanode~=~2~kV.
The measured charge gain is 2-to-5 times lower than that obtained by Policarpo et al in \cite{POLICARPO1995568} with the same MSP and at a similar anode-to-cathode voltage.
However, our estimate is not derived directly from the data but inferred from the model of Eqs  \ref{eq:model}.
Therefore, this discrepancy might be attributed to the same factors outlined above.

The dependence of the light yield with \vback showed an unexpected saturation behaviour.
This is in disagreement with the predictions obtained from finite-element computations in COMSOL and the EL \& CM model of Eqs. \ref{eq:model}.
In this calculation, the light yield grows monotonically with \vback, with a slight exponential increase above \vback~=~2~kV.
This might be explained by the nature of the the glass substrate, which may have some conductivity.
Therefore, sequential measurements at different values of \vback, might have occurred in a non-stationary field which was not included in the computation.
Further studies are required for better understanding of this observation.

The best measured energy resolution was $\sigma/\mu \approx $10\%.
This is worse than that obtained by dual-phase detectors \cite{XENON:2017lvq} in gas under a uniform field and the one reported in \cite{Aprile:2014ELthreshold} for thin wires in the liquid.
However, it is important to notice that the light collection efficiency in our setup ($\sim$9.5\%) was over a factor of 2 lower than that of \cite{Aprile:2014ELthreshold} ($\sim$21\%).
A similar light collection efficiency would improve the energy resolution to $\sigma/\mu \approx 6.5$\%, assuming that photoelectron statistics is the dominant contribution.
This value is similar to that measured in \cite{Aprile:2014ELthreshold} (7\%).
Moreover, the optimization of the field configuration by using narrower anode strips and other MSP-electrode configurations is expected to result in superior performance, as discussed in Section \ref{sec:simulations} and in \cite{Breskin:2022novel}.

The theoretical models and field calculations allowed the comparison of different MSP geometries and permitted to predict considerably enhanced light yields in LXe with the COCA-COLA and VCC geometries, as both configurations are expected to have considerably extended discharge limits.

The model-estimated light yield of such MSP devices is comparable or higher to that of dual-phase technologies, reaching values of hundreds or thousands of photons per electron, depending on the multiplier configuration and the applied potentials.
These values should be contrasted with the aforementioned yield in LXe using thin wires \cite{Aprile:2014ELthreshold, Kuger:2022} and the ones reported by, for example, the XENON collaboration using a dual-phase configuration \cite{XENON:2017lvq} and by an LHM-based detector \cite{Erdal:2018} reaching $\sim$400 photons/electron.

Notwithstanding, light production in MSPs is accompanied by charge avalanches, which may enhance the absolute light emission yields, but deteriorate the energy resolution.
Furthermore, the field configuration on the COCA-COLA and VCC designs generates electron trajectories that are vastly different, which can worsen the energy resolution of the device.
Therefore, this matter requires validation by additional investigations.

\begin{figure}
    \centering
    \includegraphics[height=72mm]{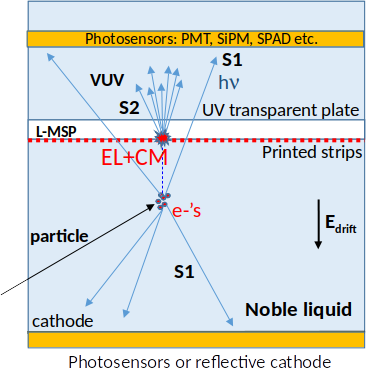}
    \includegraphics[height=72mm]{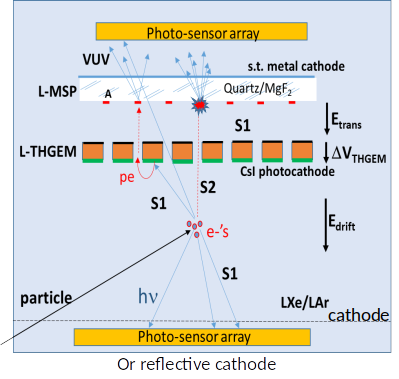}
    \includegraphics[height=72mm]{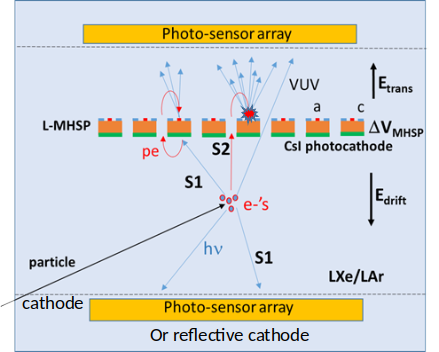}
    \caption{Potential single-phase noble-liquid detector concepts. Top left: with charge multiplication on micro-strips formed on a UV-transparent plate (e.g. MSP); emitted photons are detected on nearby photo-sensors; Top right: a CsI-coated perforated electrode followed by a MSP, permits detection of both scintillation-induced photoelectrons and ionization electrons. Bottom: a CsI-coated perforated electrode patterned on its top face with a microstrip structure \cite{Breskin:2022novel}.}
    \label{fig:concepts}
\end{figure}

Overall, our results encourage further development of MSP-based single-phase detectors.
Recently, some novel single-phase detector concepts for the detection of ionization electrons and VUV-scintillation photons were proposed in \cite{Breskin:2022novel}.
A selection of them is depicted in Figure \ref{fig:concepts}.
A key feature of these devices compared to the thin-wire approach is their superior mechanical and electrical stability.
Furthermore, unlike wires, strips can be manufactured to thicknesses down to a few nanometers, making them partially transparent to VUV light, potentially enhancing the light collection efficiency of the device.
Further developments of the concept may pave the way towards potential applications in future single-phase noble-liquid detectors.
    
\appendix
\section{Reflectivity of the MicroStrip plate at 175 nm}
\label{app:reflectivity}
The reflectivity of the plate for xenon scintillation has been estimated using Fresnel equations for reflections from the glass and chromium strips. For glass, the refractive index was assumed to be between 1.5 and 1.6 which resulted in the reflection of less than 0.2\% of the light incident to the glass surface within 0$^{\circ}$ to 30$^{\circ}$. The refraction index of liquid xenon for its scintillation wavelength was assumed to be 1.69 \cite{Solovov:2009refrIndex}. As for chromium, the experimental complex index for $\lambda = 188$~nm of $\tilde n = (1.28+1.64i)$ \cite{Johnson:1974refrIndexChromium} was extrapolated to 175~nm. The extrapolation can be justified by the fact that the plasma frequency for chromium corresponds to the photon energy of 10.75~eV, i.e. significantly higher than the energy of the scintillation photons \cite{Rakic:1998plasmaFreq}. Therefore, a smooth behavior of both parts of the refraction index can be expected. The resulting reflectivity is then estimated to be 0.258 $\pm$ 0.015, being the uncertainty due to different extrapolation scenarios. Finally, taking into account the relative area of the plate covered with chromium (0.41), we get for the fraction of reflected light an estimate of $(10.5\pm0.6)\%$.

\acknowledgments
This work was performed at the Detector Physics Laboratory of the Weizmann Institute of Science, in the context of the DARWIN dark-matter and CERN-RD51 collaborations. It was supported in part by Fundação para a Ciência e Tecnologia through project CERN/FIS-INS/0013/2021 and by CERN-RD51 Common Fund Grant.
G.M.L. and A.R. acknowledge the personal support of Dr. L. Arazi of Ben Gurion University.
We are indebted to Mr. Y. Asher of the Weizmann Institute of Science for his technical assistance.
V.C. acknowledges the personal support of the Weizmann Institute Visiting Professor Program.

\bibliographystyle{JHEP}
\bibliography{bibliography}

\end{document}